\documentclass[%
 reprint,
 amsmath,amssymb,
 aps,
]{revtex4-2}

\usepackage{graphicx}
\usepackage{dcolumn}
\usepackage{bm}
\usepackage{lineno}  

\begin{document}
\preprint{APS/123-QED}

\title{A Minimal Stochastic Variability Model of Blazars in Turbulent Cascade}

\author{Nan Ding}
 \email{orient.dn@foxmail.com}
\author{Yunyong Tang}
\affiliation{%
 School of Physical Science and Technology, Kunming University, Kunming 650214, China}%

\author{Qiusheng Gu}
\affiliation{%
School of Astronomy and Space Science, Nanjing University, Nanjing, Jiangsu 210093, China}%

\author{Rui Xue}
\email{ruixue@zjnu.edu.cn}
\affiliation{%
Department of Physics, Zhejiang Normal University, Jinhua 321004, China}%

\author{Yongyun Chen}
\affiliation{%
College of Physics and Electronic Engineering, Qujing Normal University, Qujing 655011, China}%

\date{\today}

\begin{abstract}
In this paper, we propose a novel minimal physical model to elucidate the long-term stochastic variability of blazars. The model is built on the realistic background of magnetized plasma jets dissipating energy through a turbulent cascade process that transfers energy to small-scale structures with highly anisotropic radiation. 
The model demonstrates the ability to spontaneously generate variability features consistent with observations of blazars under uniformly random fluctuations in the underlying physical parameters.
This indicates that the model possesses self-similarity across multiple time scales, providing a natural explanation for the universal power spectral density (PSD) structure observed in different types of blazars. 
Moreover, the model exhibits that when the cascade process produces a relatively flat blob energy distribution, the spectral index of the model-simulated PSD in the high-frequency regime will be steeper than that predicted by the Damped Random Walk (DRW) model, which is in agreement with recent observations of active galactic nucleus (AGN) variability, providing a plausible theoretical explanation. 
The model is also able to reproduce the observed fractional variability amplitude (FVA) characteristics of blazars, and suggests that the specific particle acceleration and radiative cooling processes within the blob may not be the key factor shaping the long-term stochastic variability. This minimal model provides a new physical perspective for understanding the long-term stochastic variability of blazars. 
\end{abstract}

\maketitle


\section{Introduction}
Blazars are a subclass of active galactic nuclei (AGN) that are among the brightest and most rapidly variable objects in the Universe. Their variability spans the entire electromagnetic spectrum from radio to gamma rays, with timescales ranging from minutes to years \cite{marscher2016variability, ghisellini2009tev, shukla2018short}. This multi-wavelength variability is one of the most prominent features of blazars and is crucial for understanding the internal physical processes governing them. 

Despite extensive observational and theoretical studies, the origin of the stochastic variability in blazars remains somewhat unclear. Some studies have focused primarily on significant short timescale events, such as flares \cite{shukla2018short, ding2019exploring,liu2019hadronuclear, shukla2020gamma}, which may be caused by shocks and/or magnetic reconnection within the jet \cite{boettcher2010timing, sironi2015relativistic, shukla2020gamma}, or by interactions of the jet with its external environment \cite{barkov2012rapid, zacharias2019extended}. However, these flare events may not represent the intrinsic long-term stochastic variability of blazars. Observational studies have shown that the most characteristic feature of the long-term variability in blazars is the power spectral density (PSD) exhibiting "red noise" behavior, where the PSD follows a power-law relation with frequency, $P(f) \propto f^{-\alpha}$, with $\alpha$ typically around -2 \cite{ryan2019characteristic, goyal2021optical, yang2021gaussian, zhang2022characterizing}. However, existing theoretical models still face challenges in explaining the mechanisms behind the long-term variability. Currently, the mainstream modeling approaches for blazar variability can be broadly categorized into two main classes: phenomenological mathematical models based on stochastic processes and physical models based on electron transport equations and radiation mechanisms. The former models, such as the Damped Random Walk (DRW) and Autoregressive Moving Average (ARMA) models, have been extensively used in modeling multi-wavelength variability in AGN, allowing for the extraction of characteristic timescales and correlation parameters of the variability \cite{yang2021gaussian,zhang2022characterizing, tarnopolski2020comprehensive, aigrain2023gaussian}. However, these models often lack a direct description of the underlying physical mechanisms driving the variability. The latter models typically focus more on specific flare events, aiming to constrain physical parameters related to particle acceleration and the emission region. Although some studies have attempted to reproduce the long-term variability of blazars by varying physical parameters of the emission region, such as magnetic field strength and particle injection rates, it often requires predefining the pattern of changes in the underlying physical parameters \cite{mastichiadis2013mrk, polkas2021numerical, thiersen2022simulations}. 

In real astrophysical environments, magnetized plasma jets often have high magnetic Reynolds numbers ($R_m >> 1$), where the strong coupling between the magnetic field and plasma flow causes the kinetic energy in the jet to be efficiently transferred from large-scale turbulent eddies to small-scale eddies through a turbulent cascade process, and then dissipated through magnetic reconnection or shock processes in the numerous small-scale structures \cite{10.1098/rsta.2014.0147,10.1103/physrevlett.121.255101,bhatta201372,webb2021nature,webb2023structure}. Recent first-principles studies of the aforementioned process have shown that the majority of particles accelerated through magnetic reconnection in small-scale structures have small pitch angles. These particles have velocities that are almost aligned with the local magnetic field direction, leading to the emitted radiation exhibiting significant anisotropy \cite{10.1103/physrevlett.121.255101,10.3847/1538-4357/ab4c33, comisso2021pitch}. 
Very recently, \citet{sobacchi2023ultrafast} has proposed a novel model in this physical background to explain the ultra-fast timescale (shorter than the crossing time of supermassive black hole event horizons) gamma-ray flares observed in blazars. The model suggests that ultra-fast flares are caused by intermittent turbulent dissipation. Energy dissipates in a group of reconnection current sheets, where electrons with small pitch angles produce narrow radiation beams that are observable only when directed towards the observer. The variability timescale of each beam is determined by the light-crossing time of a single current sheet, shorter than that of the entire radiation region, resulting in observable ultra-fast gamma-ray flares. However, this study did not further explore explanations for the long-term stochastic variability characteristics of blazars in this physical background. 

Inspired by this, based on the principle of energy conservation, this paper establishes a simple variability model in the above scenario to simulate the long-term stochastic variability of blazars. Remarkably, the model demonstrates that the naturally occurring uniform randomness of the underlying physical parameters can spontaneously give rise to the long-term stochastic variability of blazars in the form of red noise, and provides a clear physical interpretation of the characteristic quantities in the PSD. The main structure of this paper is as follows: Section 2 describes the establishment of the physical model; Section 3 presents the simulation results of the model under different parameter conditions, investigating the relation between model parameters and the characteristics of the PSD and the fractional variability amplitude (FVA). The discussion and conclusion of the model are provided in Section 4 and 5. 

\section{Physical Model}
Magnetized plasma jets of blazars derive energy from black holes or accretion disks via the Blandford-Payne (BP) or Blandford-Znajek (BZ) mechanisms, and may also acquire energy through shocks in local regions \cite{blandford2019relativistic}. 
Here, assuming a constant power $\dot{\epsilon}_{\text{jet}}$ is imparted to the jet over a time interval $\Delta \mathrm{T}$, the total dynamical energy acquired by the jet during this period is given by:  
\begin{equation}
\mathrm{E}_{\text{Total}} = \dot{\epsilon}_{\text{jet}} \Delta \mathrm{T}.  
\end{equation}
Through turbulent cascade processes, this energy is transferred from large-scale structures to small-scale structures (referred to as blobs thereafter). Turbulent cascades are highly nonlinear and complex processes, and some studies based on MHD simulations have investigated the statistical properties of the triggering of small-scale magnetic reconnection energy dissipation processes in turbulent cascades. It has been found that within a certain range of energy dissipation rates, the occurrence rate $(p)$ of magnetic reconnection exhibits a simple power-law scaling with the energy dissipation rate $(\epsilon)$ \cite{zhdankin2013statistical,knizhnik2018power}. If magnetic reconnection is the sole energy dissipation channel, $\epsilon \sim E_{\text{blob}} / E_{\text{Total}}$ and $p \sim N_{\text{blob}}/N_{\text{Total}}$, then the scaling law can naturally be extended to the energy distribution function of the blobs, following a simple power-law form as: 
\begin{equation}
    \frac{d N_{\text {blob }}}{d E_{\text {blob }}} \propto E_{\text {blob }}^{-\alpha}.  
\end{equation}
Where $E_{\text{blob}}$ is the energy of the blob, $dN_{\text{blob}}$ is the number of blobs in the energy range from $E_{\text{blob}}$ to $E_{\text{blob}}+dE_{\text{blob}}$, $\alpha$ is the spectral index of the energy distribution function of the blobs, determined by the turbulent cascade process. 
For a blob with energy $E_{\text {blob }}$, a natural consideration is to distribute its energy evenly between the magnetic field energy $E_{B}$ and the kinetic energy of particles (electrons) $E_{e}$, i.e., $E_{\text {blob }}=E_{B}+E_{e}$. When the blob is assumed to have a quasi-spherical structure, based on $E_{\text {blob }} \sim 2 U_{B} V_{\text {blob }} \sim \frac{B^{2} R^{3}_{\text {blob}}}{3}$, we have: 
\begin{equation}
B \sim\left(\frac{3 E_{\text {blob }}}{R_{\text {blob }}{ }^{3}}\right)^{1 / 2}.
\end{equation} 
Here, $U_{B}=\frac{B^{2}}{8 \pi}$ is the magnetic field energy density; $V_\text{blob}$ is the volume of the blob; $R_{\text{blob}}$ is the scale of the blob; B is the magnetic induction strength inside the blob. In the subsequent light variation realization, for simplicity, we assume that the scale of each blob is uniformly randomly distributed in the range from $R_{\text {min }}$ to $R_{\text {max }}$ (i.e., satisfying the equiprobability principle) and independent of energy. Physically, $R_{\text{min}}$ reflects the minimum scale at which turbulent dissipation occurs, while $R_{\text{max}}$ represents the maximum scale. Therefore, once the scale of a blob is determined, the magnetic induction strength inside the blob is also determined according to Equation (3).

\citet{sobacchi2021synchrotron} analyzed the synchrotron self-Compton (SSC) radiation from anisotropic particles in magnetically dominated plasma jets under fast cooling conditions, providing a set of approximate calculation formulas for SSC radiation at different pitch angles. The theory considers a plasma blob with density $n_{e}$ and size $R$, where the magnetic field fluctuation in the blob is $\delta B \sim B$. Particles are accelerated by magnetic reconnection processes and inject monochromatic energy at $\gamma \sim \sigma_{e}$ into the blob for radiation. The injected particles are parallel to the guiding magnetic field, exhibiting significant anisotropy. The angle between the particle velocity and the guiding magnetic field, known as the pitch angle, is $\theta$ $(\theta<1)$. Here, $\sigma_{e}=\frac{U_{B}}{n_{e} m_{e} c^{2}}$ is the "electron magnetization parameter", with $\sigma_{e} >> 1$ in magnetically dominated plasma. Monochromatic energy injection at $\sigma_{e}$ implies that the total electron energy density in the blob is $U_{e}=\gamma n_{e} m_{e} c^{2}=\sigma_{e} n_{e} m_{e} c^{2}=U_{B}$, where the magnetic field energy and electron kinetic energy are equally distributed within the blob, consistent with our previous assumption. After monochromatic injection, electrons undergo radiation evolution to form a steady-state electron spectrum. For an individual plasma blob, once the electron magnetization parameter $\sigma_{e}$ and pitch angle $\theta$ are determined, the analytical formulas from Table 1 to Table 6 in \cite{sobacchi2021synchrotron} can be used to calculate the synchrotron radiation energy density $U_{sy}[\varepsilon]$ and self-Compton radiation energy density $U_{IC}[\varepsilon]$ at different photon energies $\varepsilon$. This allows the total radiation energy of a single blob to be calculated as:
\begin{equation}
E_{\text {emission }}=\int_{\varepsilon_{\text {min }}}^{\varepsilon_{\text {max }}}\left(U_{s y}[\varepsilon]+U_{I C}[\varepsilon]\right) V_{\text {blob }} \mathrm{d} \varepsilon.
\end{equation}
Here, $\varepsilon_{\min}$ and $\varepsilon_{\max}$ are the integration limits of the photon energy. By selecting different integration ranges, the total radiation energy of photons in different energy bands can be obtained. 

In the implementation of light variability, a simple assumption is first considered: the total radiation energy of each plasma blob is emitted at a random time $t_{\text{blob}}$ within a duration $\Delta T$, following a Gaussian pulse profile 

\begin{equation}
F(t) = A \cdot \exp \left(-\frac{(t - t_{\text{blob}})^2}{2 \sigma^2}\right).
\end{equation}
A more realistic asymmetric pulse emission will be discussed in detail in Section IV B. 
Here, $A$ is the normalization factor determined by the total emission energy, given by $A = E_{\text{emission}} / \int \exp \left(-\frac{(t - t_{\text{blob}})^2}{2 \sigma^2}\right) dt = \frac{E_{\text{emission}}}{\sqrt{2 \pi} \sigma}$. $\sigma$ represents the standard deviation of the Gaussian pulse profile, related to the full width at half maximum (FWHM) as $\sigma = t_{\text{FWHM}} / 2 \sqrt{2 \ln 2}$. A reasonable assumption is that the radiation pulse duration of the blob is $t_{\text{FWHM}} \sim R_{\text{blob}} / c$. Thus, the light curve function of a single blob is determined by the blob energy $E_{\text{blob}}$, scale $R_{\text{blob}}$, peak time of the radiation pulse $t_{\text{blob}}$, electron magnetization parameter $\sigma_{e}$, and pitch angle $\theta$. 

Based on formulas (1) and (2), given $\dot{\epsilon}_{\text{jet}}, \Delta T, \alpha$, the total number of radiation blobs within the energy range $E_{\text{blob}} + \Delta E_{\text{blob}}$ is $\Delta N_{\text{blob}} = C \cdot E_{\text{blob}}^{-\alpha} \Delta E_{\text{blob}}$, where $C$ is a constant factor determined by the following equation:
\begin{equation}
C = \frac{\dot{\epsilon}_{\text{jet}} \Delta T}{\int_{E_{\text{min}}}^{E_{\text{max}}} E_{\text{blob}}^{-\alpha+1} dE_{\text{blob}}}
\end{equation}
Here, \(E_{\text{min}}\) and \(E_{\text{max}}\) are the minimum and maximum dissipation energies, respectively. In subsequent implementations of light variability, we use \(\left(E_{\text{min}}^{\text{blob}}, E_{\text{max}}^{\text{blob}}\right) = \left(10^{-8}, 10^{-2}\right) \dot{\epsilon}_{\text{jet}} \Delta T\), indicating that the energy dissipation rate ranges from \(10^{-8}\) to \(10^{-2}\). The model assumes that each blob is independent, and the peak radiation pulse moment \( t_{\text{blob}} \) for each blob is a uniformly distributed random number within the range \([0, \Delta T]\). This conforms to the random and intermittent dissipation of turbulence energy over the time \(\Delta T\). For simplicity, the magnetization parameter \( \sigma_e \) and the pitch angle \( \theta \) are also assumed to be uniformly distributed random numbers within the ranges \([\sigma_{\text{e,min}}, \sigma_{\text{e,max}}]\) and \([0, 1]\), respectively. 
Observationally, the ratio of the peak energy of IC (Inverse Compton) radiation to the peak energy of synchrotron radiation in blazars is approximately \(E_{IC, pk} / E_{Sy, pk} \sim 10^{6}-10^{10}\) \cite{ding2017physical,ghisellini2016blazar}. Assuming that the IC radiation occurs in the Thomson Regime, \(E_{IC, pk} / E_{Sy, pk} \sim \sigma_{e}^{2}\). Hence, in subsequent implementations of light variability, we set \(\left[\sigma_{\text{e,min}}, \sigma_{\text{e,max}}\right] = \left[10^{3}, 10^{5}\right]\) to match typical observational results. 
Up to this point, once a set of parameters \(\left(\alpha, R_{\text{min}}, R_{\text{max}}\right)\) that simply describe the jet cascading process has been established, the observed variability in jet radiation will be formed by the random and intermittent superposition of light variation pulses from a vast number of blobs: 
\begin{equation}
F_{\text{obs}}(t) = \sum_{i=1}^{N_{\text{blob}}} F^{i}(t) \cdot H\left(\theta_{i}, \boldsymbol{\beta}_{i}\right).
\end{equation}
Here, \( F^{i}(t) \) represents the light variation pulse produced by the \( i \)-th blob, and \( H(\theta_{i}, \beta_{i}) \) is a discriminant function, defined as follows:
\begin{equation}
H(\theta_{i}, \beta_{i})=\left\{\begin{array}{l}
1, \theta_{i} \geq 2 \beta_{i} \\
0, \text { others }
\end{array}\right..
\end{equation}
Here, \(\theta_{i}\) is the pitch angle of the \(i\)-th blob, which essentially is also the opening angle of the radiation produced by the blob; \(\beta_{i}\) is the angle between the radiation direction of the \(i\)-th blob and the line of sight of observation. Since the blobs are randomly distributed in the jet, \(\beta_{i}\) is a uniformly distributed random number within \([0, 2\pi]\). \(H(\theta_{i}, \beta_{i})\) reflects that we can only see those blobs whose radiation beams are directed towards the line of sight (i.e., the lighthouse effect produced by anisotropic radiation). 
\section{Simulation Results}
Based on Monte Carlo simulations, this section investigates the characteristics of the power spectral density function and the fractional variability amplitude distribution of the light curves generated by the aforementioned physical model under different parameter conditions. After specifying the constant injection power of the jet, $\dot{\epsilon}_{\text{jet}}$, and the duration $\Delta \mathrm{T}$, only three phenomenological parameters describing the cascading process in jets are needed: the spectral index of the energy distribution function of the blobs $\alpha$, the minimum dissipation scale of the blobs $R_{\text{min}}$, and the maximum dissipation scale of the blobs $R_{\text{max}}$. Using these, the random light curve of the jet within $\Delta \mathrm{T}$ can ultimately be simulated using Equation (7). 
The injection power and duration solely determine the average luminosity of the source and the duration of the variability, respectively, and do not affect the characteristics of the PSD function and the FVA distribution. Therefore, in subsequent simulations, we set $\dot{\epsilon}_{\text{jet}}=10^{45} \mathrm{erg} / \mathrm{s}$ and $\Delta \mathrm{T}=1500$ days. 
To compare with observation results, please keep in mind that the above parameters and all subsequent appearing physical quantities are in the observational frame. In contrast, the physical quantities described in the previous section are in the comoving frame of the jet. Therefore, in specific calculations, it is necessary first to transform the quantities involving time ($t$) and photon energy ($\varepsilon$) into the comoving frame for calculations, and then transform the results back into the observational frame. The transformation relations are $t_{\mathrm{obs}} = t_{\mathrm{jet}}/\delta$ and $\varepsilon_{\mathrm{obs}} = \delta\varepsilon_{\mathrm{jet}}$, where $\delta$ is the Doppler factor of the jet. In calculations, we take the typical value for blazars, $\delta = 10$. 
Figure 1(a) provides an example of a randomly generated light curve in the X-ray band $(0.1-10 \text{ keV})$ under the parameters $\alpha=1.8$, $R_{\text{min}}=1 \text{ day} \times c$, and $R_{\text{max}}=100 \text{ days} \times c$. Here, \( c \) represents the speed of light, and from $\mathrm{t}_{\mathrm{FWHM}} \sim R_{\text{blob}} / c$, it is evident that the preceding coefficients characterize the shortest/longest variability timescales of the blobs. The simulated light curve is similar to those observed in blazars, featuring a rich structure, including extremely rapid short-term flares, long-term asymmetric outbursts, and smaller flares superimposed on these.
\subsection{The simulation results of the PSD}
\begin{figure*}
  \centering
    \includegraphics[width=0.45\textwidth]{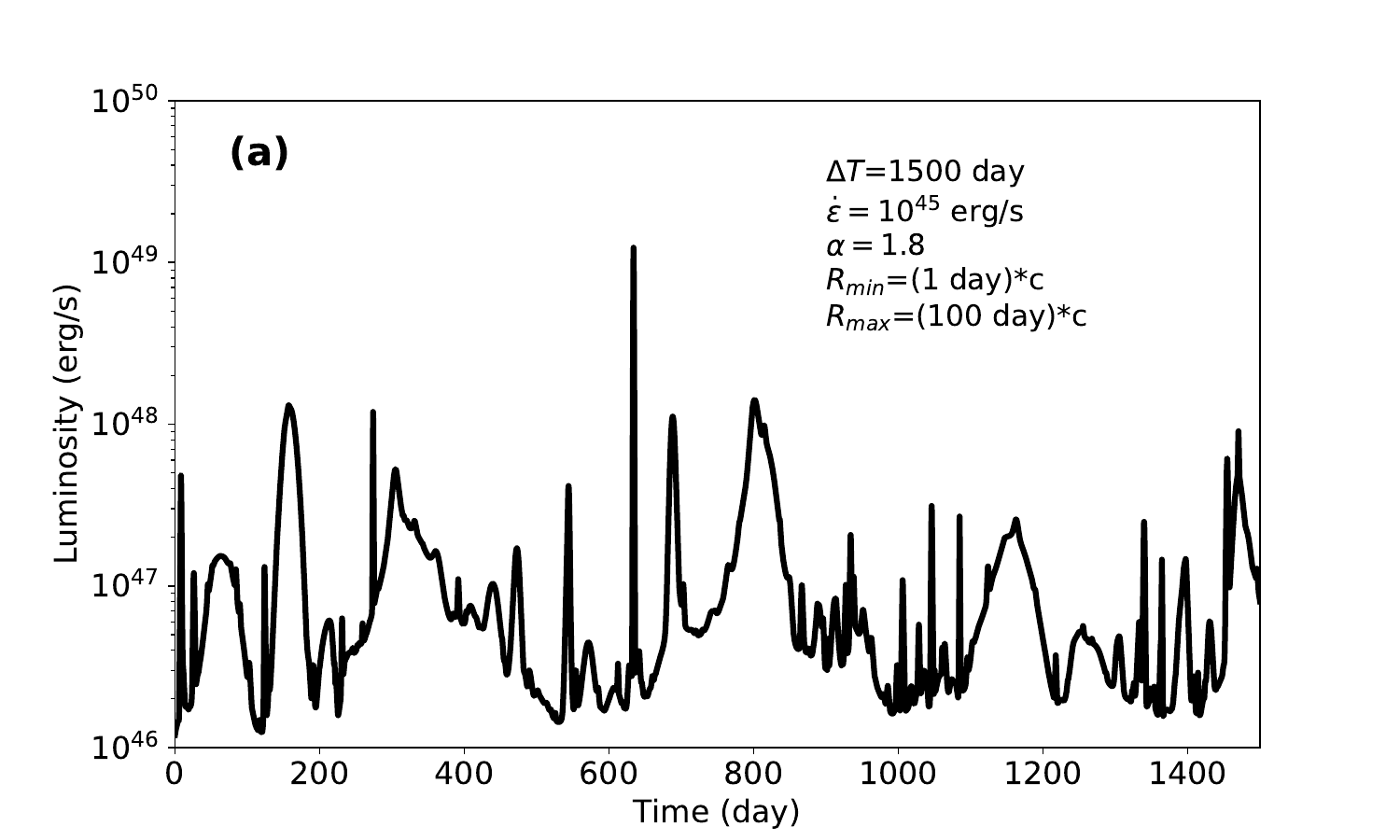}
    \includegraphics[width=0.45\textwidth]{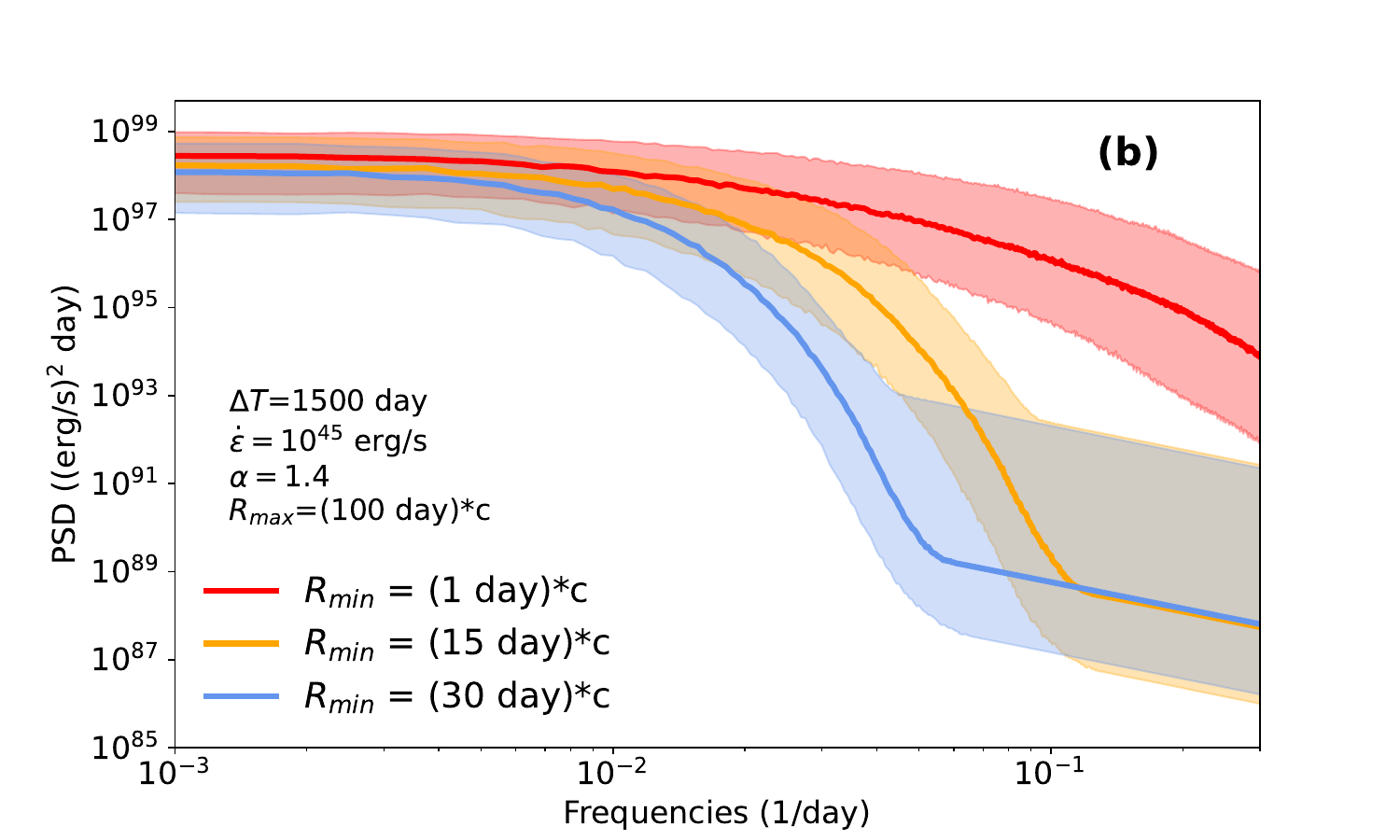}
    \includegraphics[width=0.45\textwidth]{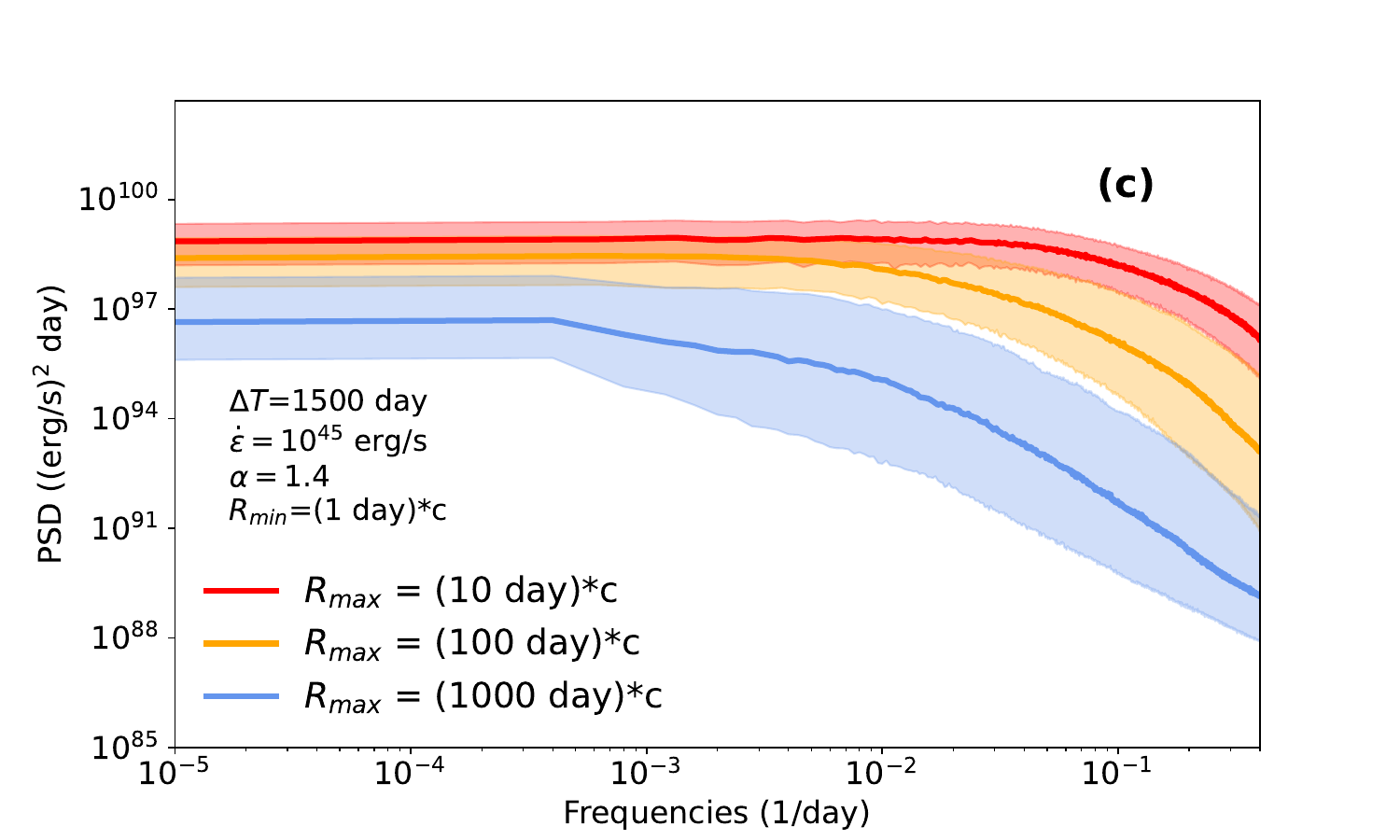}
    \includegraphics[width=0.45\textwidth]{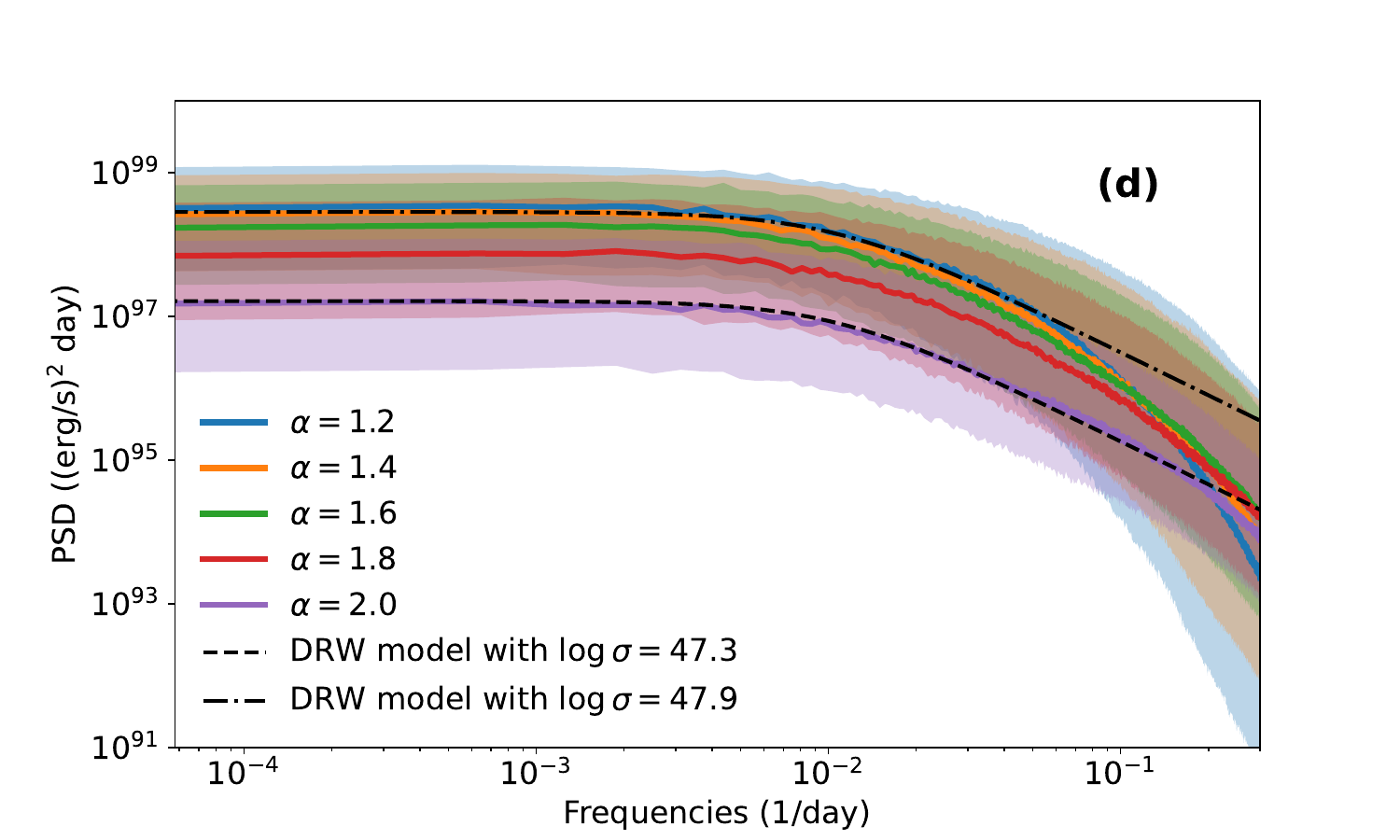}
\caption{A simulated light curve from a Monte Carlo simulation based on our model (a) and the ensemble power spectral density function of the light curves generated under different parameter conditions. The corresponding model parameters are displayed on each panel. (b) shows the results for variable $R_{\min}$ cases; (c) shows the results with $R_{\max}$ as the variable parameter; (d) shows the results with $\alpha$ as the variable parameter. The error bands represent the 1$\sigma$ uncertainty. See Section 3.1 for more details. }
\end{figure*}
Figure 1(b) presents the ensemble PSD functions of the X-ray light curves simulated for different values of $R_{\text{min}}$ when $\alpha=1.4$ and $R_{\text{max}}=100~\mathrm{day} \times c$. 
Specifically, under a given set of model parameters, 5000 random light curves are simulated, and the PSD is calculated for each light curve, ultimately obtaining the ensemble PSD function for that set of parameters. The Power Spectral Density (PSD) of the simulated light curves exhibits a double break power-law form. Below the low-frequency break, the PSD is independent of frequency, showing a plateau; between the low and high-frequency breaks, the PSD approximately follows a power-law distribution; above the high-frequency break, the PSD no longer changes significantly with frequency, approaching a flat spectrum.
The parameter $ R_{\min} $ primarily determines the high-frequency break position of the PSD, and due to the limitation on the break point, it can lead to changes in the slope of the PSD. Current observational studies indicate that the PSD of high-energy radiation variability in blazars does not show a clear high-frequency break within the frequency range of $ 10^{-5}-10^{-1} $ 1/day; the PSD exhibits a single break power-law form (as shown by the red curve in Fig (b)), with the break occurring in the low-frequency region (break point $ \nu_{\text{br}} \sim 10^{-2}-10^{-3} $ 1/day) \cite{tarnopolski2020comprehensive,bhatta2020nature,yang2021gaussian,zhang2022characterizing, goyal2022multiwavelength}. Above the low-frequency break, the PSD follows a power-law spectrum with $ \alpha_{\text{PSD}} \sim 1-2 $ (PSD $ \propto \nu^{-\alpha_{\text{PSD}}} $); below the break point, the PSD flattens. Within this model framework, the observed results imply that the minimum scale of turbulent cascade dissipation, $ R_{\min} \ll 15 ~ \text{days} \times c \sim 10^{16}$ cm. Dissipation at these small scales would produce the rapid flares observed in blazars, occurring on timescales of hours or even minutes. 

Figure~1(c) presents the ensemble PSD function of the X-ray light curves simulated with different parameters for $\alpha=1.4$, $R_{\text{min}}=1~\text{day} \times c$, and varying $R_{\text{max}}$. 
It can be seen that in the frequency range primarily covered by current observations, with a smaller $R_{\text{min}}$, the PSD exhibits a single-break power-law form. The parameter $R_{\text{max}}$ mainly affects the low-frequency break position of the PSD. $R_{\text{max}}$ has minimal influence on the PSD slope in the high-frequency region, with the spectral index in this region being essentially around 2. 
The overall characteristics of the PSD are similar to those produced by a damped random walk (DRW) stochastic process. DRW is a stochastic process where the current value is a combination of the previous value and random fluctuations. Mathematically, DRW is a special case of a first-order continuous autoregressive process, CAR(1). The PSD of DRW is as follows\cite{kelly2009variations}:
\begin{equation}
\operatorname{PSD}(f)=\frac{2 \sigma^{2} \tau^{2}}{1+(2 \pi f \tau)^{2}}.
\end{equation}
Here, $\tau$ represents the relaxation time of the CAR(1) process, often referred to as the characteristic timescale in the DRW model, while $\sigma$ reflects the intensity of variability. Therefore, the DRW in the high-frequency regime behaves as a $\operatorname{PSD}(f) \propto f^{-2}$ process, similar to an ordinary random walk; at low frequencies $(f \ll 1 / \tau)$, the "damping" characteristics are observed, manifesting as a flat PSD. 
In recent years, observational studies have shown that the DRW model has been successful in quantitatively describing the variability of both radio-quiet and radio-loud AGNs. However, a limitation is that the DRW model is essentially a phenomenological mathematical model, lacking physical meaning. The characteristic timescale $\tau$ in the model is speculated to be related to some physical quantity within the system (for example, for radio-quiet AGNs, it is speculated to be related to the mass of the black hole \cite{papadakis2004scaling}; for radio-loud AGNs, it might be related to the thermal instability timescale of the accretion disk \cite{zhang2022characterizing}). 
Under the current model, the uniformly random model parameters, adhering to the principle of equi-probability, naturally generate a PSD similar to that characteristic of the DRW process. In this scenario, the characteristic timescale of the DRW model, i.e., the low-frequency break point, is determined by the maximum scale of energy dissipation in the turbulent cascade. The observational studies of Blazars in X-ray and gamma-ray show that the break frequency is $\nu_{br} \sim 10^{-2} - 10^{-3}$ 1/day\cite{bhatta2020nature,yang2021gaussian,zhang2022characterizing}. Within the current model framework, this suggests that the maximum energy dissipation scale of the turbulence is $R_{\max} \sim (100-1000)~\text{day} \times c \sim (10^{17} - 10^{18})~$cm. 
This result is significantly larger compared to the size of a single turbulent blob assumed in some previous works using shock turbulence models to simulate blazar variability \citep[][]{bhatta201372,webb2021nature,webb2023structure}, which appears to be a contradiction. 
However, it should be noted that the timescales simulated in these works for variability range only from a few days to tens of days (i.e., micro-variability). 
Here, to be in line with the observational results of the low-frequency break in the PSD (corresponding to variability timescales on the order of years), it is necessary to have large-scale energy dissipation structures with variability timescales ranging from months to years, which contribute to the long timescale background variation (distinct from micro-variability). 
Therefore, $R_{\max}$ cannot be simply understood here as the maximum scale of a single turbulent cell, but its size reflects the maximum scale that the entire turbulent dissipation region could potentially reach. 
From \cite{webb2021nature}, it is known that the maximum scale of a single turbulent cell can reach $\sim 100$~AU, and the maximum number of cells within the turbulent region can reach $\sim 100$. This implies that the maximum scale of the entire turbulent dissipation region can indeed reach $\sim 10^4$~AU ($10^{17}$~cm). 

Figure~1(d) presents the ensemble PSD functions of the simulated X-ray light curves under different parameter settings of $\alpha$ with $\mathrm{R}_{\min}=1~\text{day} \times c$ and $\mathrm{R}_{\max}=100~\text{day} \times c$. 
For comparison, the PSD functions of the DRW model with peak values comparable to $\alpha=1.2$ and 2.0 are also shown (black dotted and dashed lines). 
The results indicate that the choice of $\alpha$ has little impact on the high/low break frequencies of the PSD functions, but it slightly modulates the peak PSD values, which decrease as $\alpha$ increases. Compared to the DRW model, this model demonstrates consistent PSD results in the transitional frequency range from the low-frequency plateau to the high-frequency power-law form. However, as the frequency increases further, the PSD spectral index becomes steeper, deviating from the DRW model. Noticeably, as the parameter $\alpha$ decreases, this deviation becomes more pronounced. For instance, when $\alpha=1.2$ (blue line), the PSD starts to significantly deviate from the DRW model at frequencies $f>0.06$~$\text{day}^{-1}$. In contrast, for $\alpha=2.0$ (purple line), the PSD only exhibits a slight deviation at $f>0.1~\text{day}^{-1}$. This feature reveals a clear difference between the current model and the DRW model. Interestingly, recent observational results on quasars have revealed features consistent with the current model predictions. \citet{mushotzky2011kepler} found that quasar light curves measured by Kepler exhibit PSDs with a slope of -2.6 to -3.3 on very short timescales (around a few days), deviating from the -2 expected under the DRW model. \citet{zu2013quasar}, based on a sample of OGLE quasar light curves, considered four modified covariance functions to search for evidence of quasar variability deviating from the DRW model. They also found that while the DRW model generally describes the variability well on long timescales, there are indications of deviations from the DRW model on very short timescales. \citet{stone2022optical} studied the optical gri-band variations of a sample of 190 quasars in the SDSS Stripe 82 region, confirming that the ensemble PSD of optical variability on monthly timescales is steeper than the DRW model prediction. However, it should be noted that most of these studies have focused on radio-quiet AGN, and mainly on the optical band. Relevant research and reports on the high-energy variability of Blazars deviating from the DRW model are relatively scarce. The current research on the PSD of high-energy variability of blazars is primarily focused on the gamma-ray band, mainly based on the Fermi-LAT data, and the observed PSD frequency range is typically in the range of $10^{-1}$ to $10^{-4}$ day$^{-1}$ \cite{yang2021gaussian,zhang2022characterizing,tarnopolski2020comprehensive}. Within this range, the PSD of blazars follows the DRW model, with no obvious deviations (although to some extent limited by observational uncertainties). This observational result indicates that in the current model, the spectral index $\alpha$ is constrained to $\alpha \gg 1.2$. 

\subsection{The simulated results of the FVA}
\begin{figure*}
  \centering
    \includegraphics[width=0.49\textwidth]{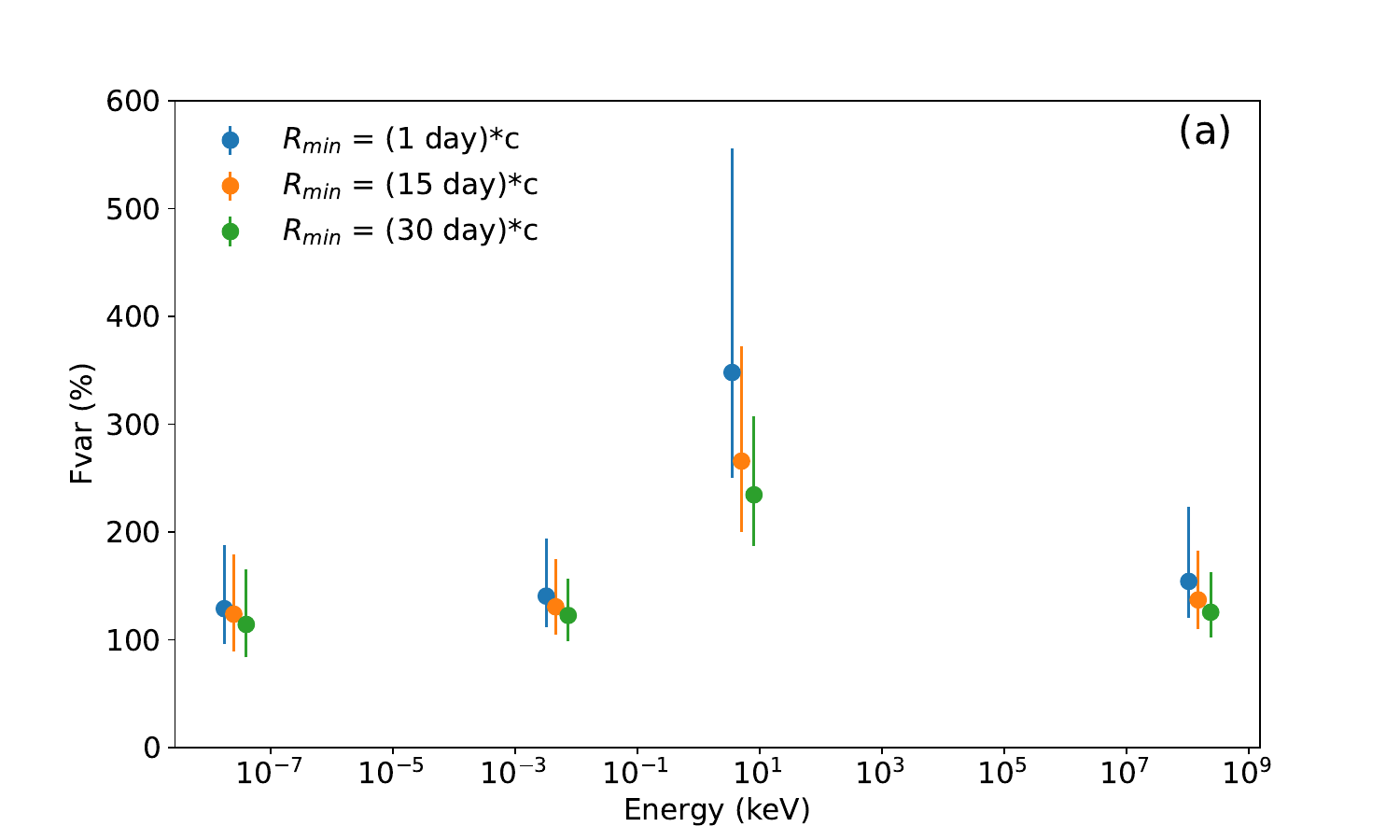}
    \includegraphics[width=0.49\textwidth]{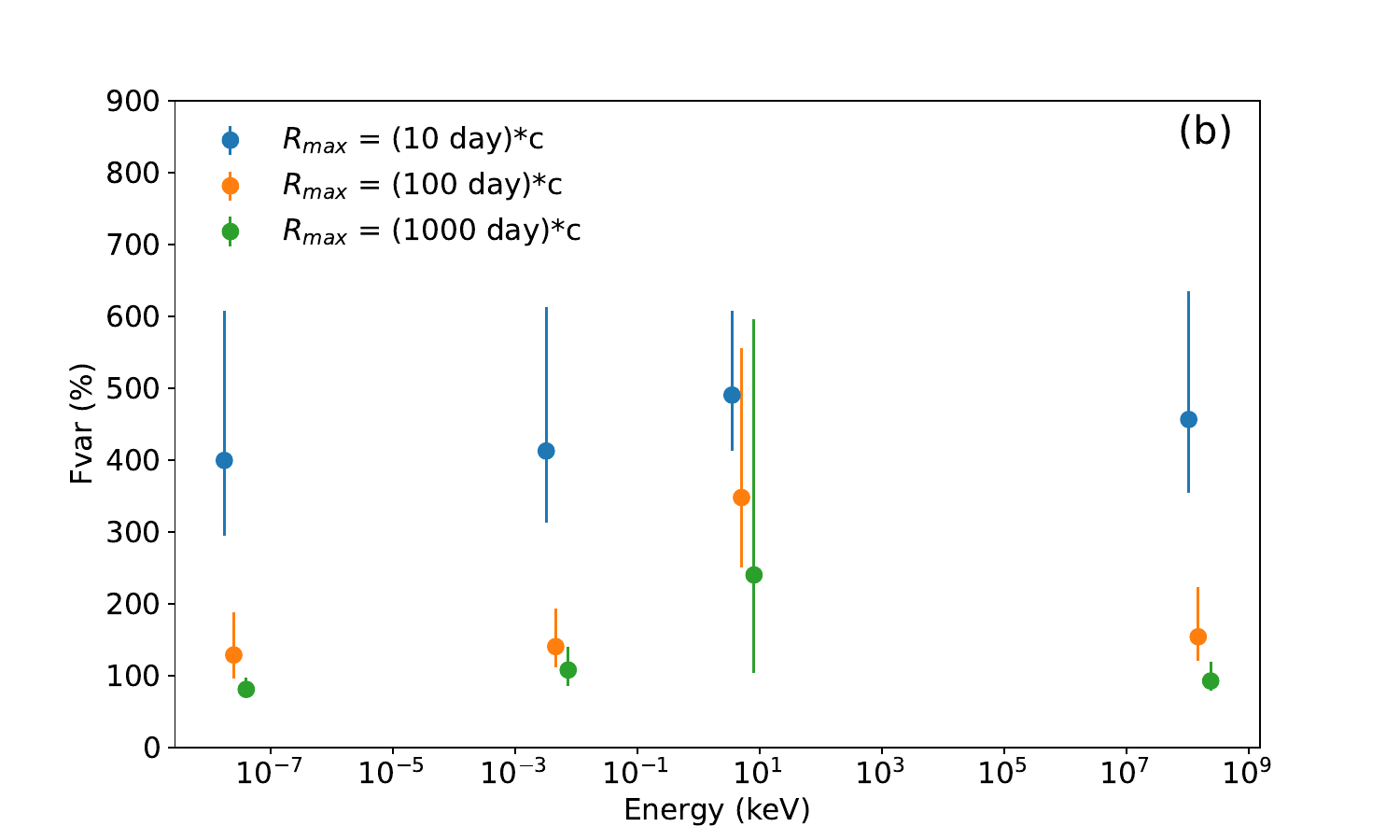}
    \includegraphics[width=0.49\textwidth]{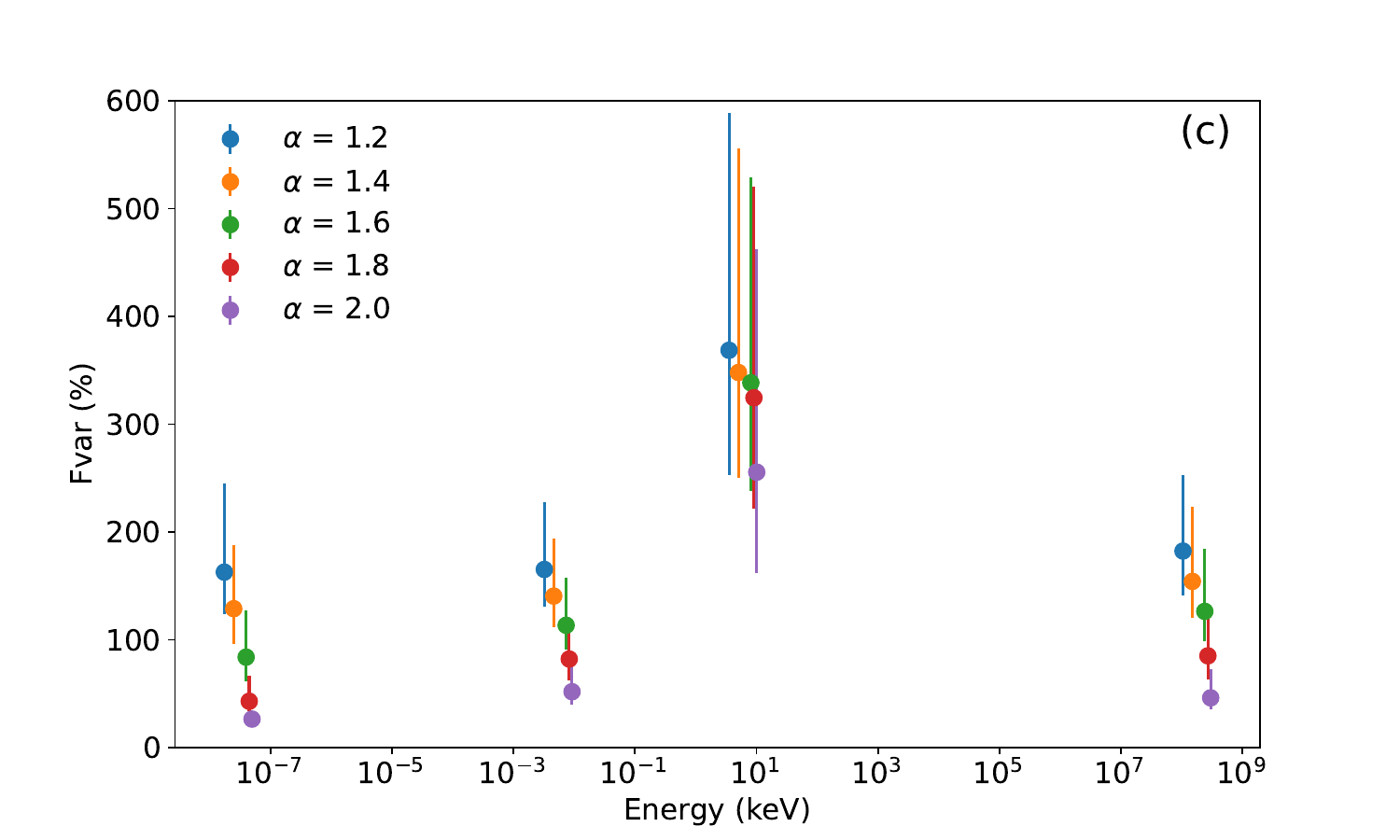}
\caption{The fractional variability amplitude spectrum of the light curves generated by the model under different parameter conditions. The fixed model parameters correspond to those in Figure~1. To avoid overlapping data points, the data points obtained from each parameter set are slightly offset in the horizontal direction. (a) Results for variable $R_{\min}$; (b) Results for variable $R_{\min}$; (c) Results for variable $\alpha$. The error bands represent the 1$\sigma$ uncertainty. Refer to Section 3.2 for details.} 
\end{figure*}
In observational studies, the fractional variability amplitude ($F_{\text{var}}$) is a commonly used dimensionless parameter to characterize the strength of flux variability. It is defined as:
\begin{equation}
F_{\mathrm{var}} = \sqrt{\frac{S^2 - \langle \sigma_{\text{err}}^2 \rangle}{\langle F \rangle ^{2}}}.
\end{equation}
Here, $S^2$ is the variance of the light curve data, $\langle \sigma_{\text{err}}^2 \rangle$ is the average variance of the measurement errors in the data, and $\langle F \rangle$ is the mean flux. Based on the simulation results, we calculated the FVA spectra under different parameter conditions. Specifically, based on the emission spectrum of the source, we obtained the light curves in the radio ($4-8$ GHz), optical ($170-650$ nm), X-ray ($0.1-10$ keV), and gamma-ray ($0.1-300$ GeV) bands by integrating the radiation photon energy over different upper and lower limits. Then, based on Equation (11) and taking $\langle \sigma_{\text{err}}^2 \rangle=0$, we calculated the intrinsic FVA of 5000 randomly generated light curves for each parameter set and each band. Finally, the ensemble FVA spectra were compiled. 

Figure~2 shows the simulation results when $R_\mathrm{min}$, $R_\mathrm{max}$, and $\alpha$ are used as independent variables, respectively. The values of the control variables are consistent with the implementation corresponding to Figure~1. The parameters $R_\mathrm{max}$ and $\alpha$ greatly affect $F_\mathrm{var}$, while $R_\mathrm{min}$ has a small impact. $F_\mathrm{var}$ is inversely related to $R_\mathrm{min}$, $R_\mathrm{max}$, and $\alpha$, decreasing as they increase. 
Additionally, it is observed that the variability amplitude in the X-ray band is the highest among all bands, surpassing the low-energy band by approximately 1-3 times. This feature is generally consistent with the observational results of FVA spectra of several blazars (e.g., Mrk 421 \cite{aleksic20152009} and Mrk 501 \cite{ahnen2017multiband}). 
\citet{richards2014connecting} conducted a statistical analysis of 4-year 15 GHz radio light curves of 1500 blazars observed with the 40-meter radio telescope at Owens Valley Radio Observatory, and found that the average $F_\mathrm{var}$ of these blazars is around 20\%, mainly concentrated in the range of 10\% to 30\%. \citet{zhang2015long} used the optical monitoring data from the Small and Moderate Aperture Research Telescope System (SMARTS) to calculate the FVA in the R and J bands of 49 flat-spectrum radio quasars (FSRQs) and 22 BL Lacertae objects (BL Lacs), and found that the average $F_\mathrm{var}$ is around 30\%, mainly concentrated in the range of 20\% to 70\%. By comparing these observational results and simulation results, we can constrain $R_{\max} > 100~\text{day} \times c$ and $\alpha > 1.6$, which are consistent with the results obtained based on the PSD. 

In this model, the energy of each emitting blob is assumed to be released in the form of a Gaussian pulse profile within a certain time width, without strictly accounting for the specific acceleration and radiative cooling process of particles within the blob.
Nevertheless, the simulation results are able to essentially reproduce the observational characteristics of the PSD and FVA spectra of blazars. This suggests that the specific dynamical processes within individual blobs during flaring are not a critical factor in the formation of the red noise characteristic long-term stochastic variability. In fact, the universality exhibited by different types and luminosities of blazars in their PSD and FVA spectra suggests that the physical mechanism driving the variability should possess self-similarity across multiple timescales. 

\section{Discussion}
\subsection{Analytical Analysis of the Model}
The model proposed here is based on the highly anisotropic radiation emitted by small-scale structures formed through turbulent cascading processes. Essentially, the stochastic variability of blazars is attributed to the superposition of a large number of discrete flare events. Interestingly, this model can spontaneously generate red noise-like variability behavior, even though the underlying physical parameters are all (naturally) uniformly random. To more intuitively reveal how the red noise-like PSD structure arises from the statistical properties of the superposition of a large number of discrete flare events, here we conducted further analytical analysis and discussion of this model. 

The light curve of an individual blob in the model is assumed to follow the Gaussian pulse profile as described by Equation (5). In the frequency domain, the PSD function of this pulse is $\operatorname{PSD}(\omega) \propto |\text{FT}[F(t)]|^2 = A^2 \sigma^2 \exp(-\sigma^2 \omega^2)$, which is independent of the peak time $t_{\text{blob}}$ of the flare events. 
Therefore, the PSD function of the final light curve (i.e., equation (7)) formed by the superposition of a large number of random, intermittent light pulses is: 
\begin{equation}
\begin{aligned}
  \operatorname{PSD}(\omega) \propto \sum_{i=1}^{i=N_{\text{obs}}} A_{i}^{2} \sigma_{i}^{2} \exp \left(-\sigma_{i}^{2} \omega^{2}\right) \\ = \sum_{i=1}^{i=N_{\text{obs}}} \frac{E_{\text {emission}_i}^{2}}{2 \pi} \exp \left(-\sigma_{i}^{2} \omega^{2}\right).  
\end{aligned}
\end{equation}
Here, $N_{\text{obs}}$ is the number of observed blobs along the line of sight. 
Considering that $E_{\text{emission}_i}\propto U_{\mathrm{B}_i}V_{\text{blob}_i} \propto E_{\text{blob}_i}$ and the approximation $\exp\left(-\sigma_i^2 \omega^2\right) \approx \frac{1}{1+\sigma_i^2 \omega^2} + \mathcal{O}\left[\sigma_i^4 \omega_i^4\right]$, it follows that:

\begin{equation}
\operatorname{PSD}(\omega) \propto \sum_{i=1}^{i=N_{\text{obs}}}\left(\frac{E_{\text{blob}_i}^2}{1+\sigma_i^2 \omega^2} + \mathcal{O}\left[\sigma_i^4 \omega_i^4\right]\right).
\end{equation}

When $\sigma_{i}^{4} \omega^{4} \ll 1$, the expression can be approximated as the superposition of $N_{\text{obs}}$ DRW processes with different variability amplitudes and characteristic timescales (i.e., Equation (9)). 
At this point, when $\sigma_{i}^{2} \omega^{2} \ll 1$, consistent with a single DRW process, $\operatorname{PSD}(\omega)\propto\sum_{i=1}^{i=N_{\text{obs}}} E_{\text{blob}_{i}}^{2}$, the PSD will exhibit a "damped" characteristic, appearing as a plateau in the low-frequency regime; when $\sigma_{i}^{2} \omega^{2} \leq 1$, $\operatorname{PSD}(\omega) \propto \sum_{i=1}^{i=N_{\mathrm{obs}}}\left(\frac{E_{\mathrm{blob}_i}^{2}}{1+\sigma_{i}^{2} \omega^{2}}\right)$. In the limit where the number of blobs, $N_{\text{blob}}$, is large, under the model assumptions, the energy of individual blobs, $E_{\text{blob}_i}$, follows a power-law distribution with spectral index $\alpha$, and the variance $\sigma_i$ of individual blobs follows a uniform distribution between $\sigma_{\min}$ and $\sigma_{\max}$. In this case, the PSD can be further expressed as:

\begin{equation}
\resizebox{0.5\textwidth}{!}{$
\begin{aligned}
\operatorname{PSD}(\omega) &\propto \int_{E_{\min }}^{E_{\max }} C E_{\text {blob }}^{2-\alpha} \mathrm{d} E_{\text {blob }} \int_{\sigma_{\min}}^{\sigma_{\max}} \frac{1}{\sigma_{\max}-\sigma_{\min}} \frac{1}{1+\sigma^{2} \omega^{2}} \mathrm{d} \sigma \\
&=\frac{C\left(E_{\min }^{3-\alpha}-E_{\max }^{3-\alpha}\right)}{-3+\alpha} \cdot \frac{\arctan \left(\sigma_{\max}\omega \right)-\arctan \left(\sigma_{\min} \omega \right)}{(\sigma_{\max} - \sigma_{\min} ) \omega}.
\end{aligned}
$}
\end{equation}

Here, $C$ is a constant, and $E_{\min}$ and $E_{\max}$ are the minimum and maximum values of the individual blob energy, respectively.  
Given $\arctan (\sigma \omega) \approx \sigma \omega-\frac{(\sigma \omega)^{3}}{3}+\mathcal{O}\left(\sigma^{5} \omega^{5}\right)$ and $\sigma_{\max } \gg \sigma_{\min }$, it can be approximated as:  
\begin{equation}
\operatorname{PSD}(\omega) \propto \frac{C \left(-E_{\max }^{3-\alpha}+E_{\min }^{3-\alpha}\right)}{-3+\alpha} \cdot \frac{1}{1+ \frac{1}{3}\sigma_{\max }^{2} \omega^{2}}.  
\end{equation} 
In the limit as \( N_{\text{obs}} \) approaches infinity, the superposed PSD converges to the PSD of a single DRW process with a characteristic timescale of $\tau = \frac{\sigma_{\max}}{2\sqrt{3}\pi}$. This convergence explains the similarity between the model-simulated PSD and the DRW model PSD in the low-frequency regime, as well as the dependence of the PSD low-frequency break point on \( R_{\text{max}} \) (as shown in Figure 1 (c)). The condition $\sigma_{\text{max}} \gg \sigma_{\text{min}}$ implies that when $\sigma_{\text{min}}$ is relatively large, the model PSD significantly deviates from the DRW process (as seen in Figure 1(b)). Furthermore, the large-number limit condition of $N_{\text{obs}}$ suggests that as the blob energy distribution spectral index $\alpha$ decreases (resulting in a significant decrease in $N_{\text{obs}}$), the deviation of the PSD from the DRW process will become more pronounced (as seen in Figure 1(d)). 

The PSD characteristics of long-term random variability in blazars can be attributed to the statistical properties of the superposition of numerous discrete flare events. This is not surprising, as red noise PSDs are common in nature, often explained by the superposition of independent processes with suitable relaxation time distributions \cite{kaulakys2005point}. While this is a general mathematical explanation, conducting specific analyses for the actual problems remains meaningful, as demonstrated here, which assigns clear physical meanings to parameters. 

\begin{figure*}
  \centering
    \includegraphics[width=0.49\textwidth]{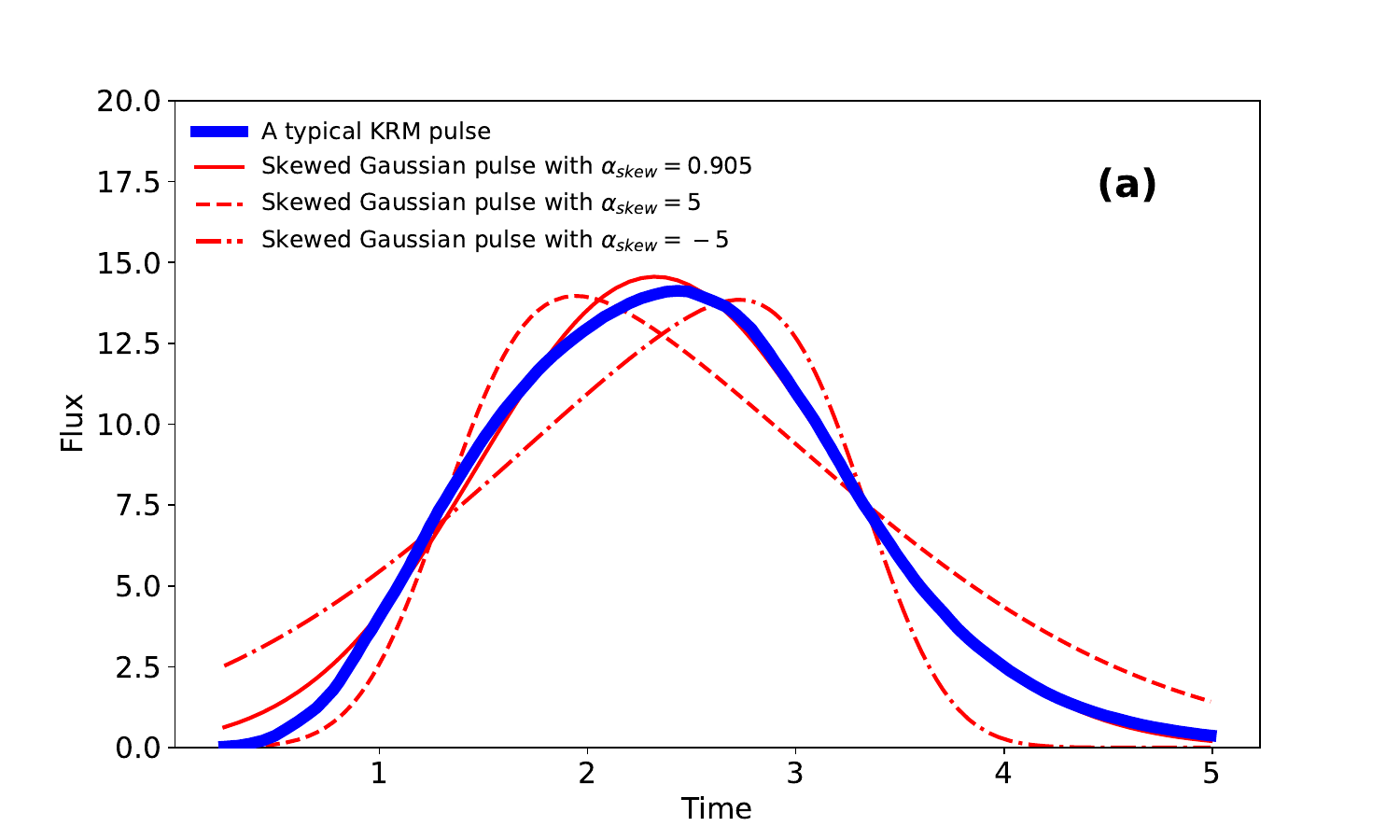}
    \includegraphics[width=0.49\textwidth]{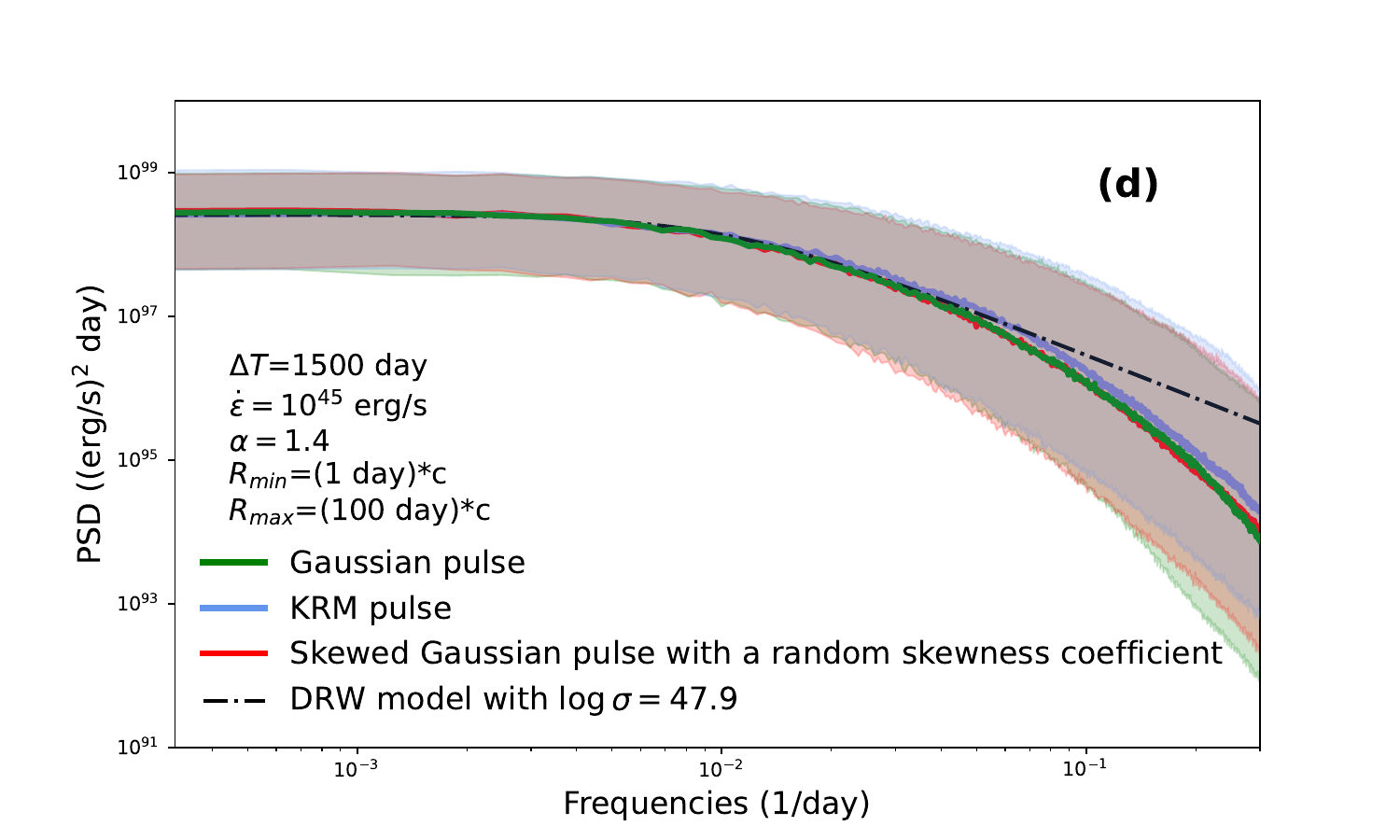}
\caption{(a) Asymmetric pulses in different forms. The blue curve is a typical KRM pulse from \cite{webb2023structure}. (b) The simulation results of the PSD under asymmetric pulse scenarios. 
The results of the PSD under the Gaussian pulse scenario and the corresponding DRW model from Figure 1(d) are also shown here for comparison. 
} 
\end{figure*}

\subsection{Impact of Asymmetric Impulses on Model Results}
In the aforementioned model, we assume that the pulse of the blob takes the form of a simple Gaussian profile. Although this assumption facilitates analytical analysis, in reality, flares driven by either magnetic reconnection or shock typically exhibit asymmetric pulse profiles \cite{kirk1998particle,shukla2020gamma,bhatta201372,webb2021nature,webb2023structure}.
For example, recently, researchers from the Florida International/SARA blazar monitoring program analyzed hundreds of complex micro-variability light curves of blazars, decomposing them into individual pulses \cite{webb2021nature}. They found that these individual pulse structures exhibit similarities, and their morphology is consistent with the pulse profiles calculated by the theoretical model in \cite{kirk1998particle}, hereafter referred to as KRM pulse. 
The KRM pulse is generated by particle acceleration and radiative cooling processes when a relativistic shock encounters a region of enhanced plasma density (such as a turbulent cell). It exhibits a distinct asymmetry. 
To examine how asymmetric pulses affect our model, we calculate PSDs for the following two scenarios with asymmetric blob radiation: 

\textit{Scenario 1:} Assume that the pulse structure of each blob follows a typical KRM pulse (i.e., the blue curve in Figure 3a). Following the method of \cite{webb2023structure}, each blob pulse is generated by adjusting the center time ($t_\text{blob}$), FWHM ($t_\text{FWHM}$), and amplitude ($A$) while maintaining its basic shape. The amplitude of the pulse is calculated from $E_\text{emission}$, using the same method as for Gaussian pulses. 
Note that this Scenario implies an assumption that the pulses of the turbulent blobs are triggered by shock, rather than magnetic reconnection. 

\textit{Scenario 2:} The pulse profile is described using the following skewed Gaussian function: 
\begin{equation}
    f(t) = A \cdot \phi \left( \frac{t - t_\text{blob}}{\omega} \right) \Phi \left( \alpha_\text{skew} \left( \frac{t - t_\text{blob}}{\omega} \right) \right).
\end{equation}
Here, $ \phi $ is the probability density function of the standard normal distribution, and $ \Phi $ is the cumulative distribution function of the standard normal distribution. 
Similar to the Gaussian function, $t_\text{blob}$ is the central time of the pulse, and $A$ is the normalization factor determined by the total radiated energy of the blob. $\omega$ is the scale parameter that controls the width of the pulse, determined by the FWHM. $\alpha_\text{skew}$ is the shape parameter used to adjust the asymmetry of the pulse. 
By adjusting this parameter, various asymmetric pulse structures can be realized. Figure 3(a) shows the skewed Gaussian pulses for $\alpha_\text{skew}$ values of -5, 0.905, and 5. Notice that when $\alpha_\text{skew} = 0.905$, the skewed Gaussian pulse is closest to the typical KRM pulse, with a maximum relative error of only $\sim 6\%$ within the FWHM range. 
Using skewed Gaussian pulses, an extreme scenario is considered where the pulse shape of each blob is different and random. This is achieved in the simulation by randomly sampling $\alpha_{skew}$ from a uniform distribution over [-5, 5]. 

Figure 3(b) presents the PSD simulation results for $\alpha=1.4$, $R_{\text{min}} = 1~\text{day} \times c$, and $R_{\text{max}} = 100~\text{day} \times c$ for the above two scenarios. For comparison, the PSD results for Gaussian pulses and the DRW model are also shown. The results demonstrate that the PSDs for scenarios 1 and 2 are in agreement with that obtained for the Gaussian pulse, with only slight discrepancies in the high-frequency region. 
This result is predictable, as on the short timescale, the profile of individual pulse is resolved, and the differences in pulse structure will leave imprints in the high-frequency region of the PSD. However, the overall result of the PSD is determined by the statistical property of the superposition of blob pulses and does not depend on the individual pulse structure, thus demonstrating a universality. Moreover, it is interesting to note that the differences in the high-frequency region for scenario 1 are relatively more pronounced. This can be attributed to the fact that the radiation from each blob has a consistent KRM pulse structure. In scenario 2, although the pulse shapes of each blob are different, the result of the superposition of these pulses will converge to the result of the Gaussian pulse according to the central limit law under the condition that the asymmetry is uniformly random. 
From the above analysis, the impact of blob asymmetric pulses on the main results of the current model is limited. 

\subsection{Comparison with Previous Models}
As mentioned in the introduction, the current theoretical models for the random variability of blazars can be broadly categorized into two main types: one is based on phenomenological mathematical models of stochastic processes, such as the DRW model and more flexible ARMA or continuous autoregressive moving average (CARMA) models. These models have been widely used in modeling the multi-band variability of AGN to extract the characteristic timescales and relevant parameters of the variability. However, they lack a direct description of the physical mechanisms behind the variability. For example, the physical significance of the low-frequency break characteristic timescale in the PSD is not clear. 
\citet{brill2022variability} proposed a novel autoregressive inverse gamma light curve model to explain the heavy-tailed flux distributions observed in blazar variability. The model suggests that the gamma-ray variability of blazars arises from the collective impact of discrete flare events driven by a Poisson process. Since individual flare events are often unresolved within observation intervals, the measured quantity in each interval is the average flux. By employing sparsification techniques, Brill developed a first-order autoregressive (AR(1)) process that conforms to an inverse gamma distribution. This process effectively replicates the random long-term light curves of blazars with only three free parameters: average burst rate, burst fluence, and autocorrelation timescale. 
While the model proposed by Brill and the turbulent cascade dissipation scenario considered here have different physical starting points, both models attribute the essence of variability in blazars to the superposition of discrete flare events. However, in our model, the physical significance of the model parameters is more explicit, with the time structure of a single flare being distinguishable, and the model not assuming a predetermined form for the underlying parameters driving the variability. The red noise characteristics of the light curve can spontaneously emerge. This property indicates that the current model exhibits self-similarity at multiple time scales, naturally explaining why different types of blazars exhibit universal PSD structure. 

The second category of models is grounded in electron transport equations and radiation mechanisms for the analysis of variability. These models predominantly concentrate on specific flare events to constrain the pertinent parameters governing acceleration and radiation processes. Certain investigations seek to replicate the random long-term variability of blazars by fine-tuning parameters like the magnetic field intensity in the emission region and the rate of particle injection \cite{mastichiadis2013mrk, polkas2021numerical, thiersen2022simulations}. For instance, in a recent study by \citet{thiersen2022simulations}, a dynamically evolving single-zone leptonic model was utilized. This model presupposes a random process with a power-law PSD to emulate the temporal fluctuations of physical parameters within the emission region (such as electron injection luminosity, magnetic field intensity, and electron injection spectral index) to investigate the distinctive imprints of diverse radiation mechanisms and variable parameters on the PSD characteristics and time-delay correlations of light curves. 
\citet{finke2014fourier} developed a theoretical model that, by analyzing electron transport equation and radiation processes in the Fourier domain, enables direct comparison with the observed PSD. The model considers that the only physical quantity in blazars that varies with time is the rate of electron injection into the radiation region. To replicate the observed red noise variability characteristics, the model necessitates a power-law relation between the electron injection rate and the injection frequency (i.e., injection time). 
\citet{marscher2013turbulent} creatively proposed a Turbulent Extreme Multi-Zone (TEMZ) model to simulate flux and polarization variations in blazar jets. In this model, relativistic plasma flows through the jet and passes through a stationary conical shock. The shock compresses the plasma and accelerates electrons to high-energy states. Turbulence in the model is approximated as a large number of blobs with random uniform magnetic fields, and the superposition of radiation from these blobs ultimately produces the light curve. This aspect is similar to our model. However, in terms of the realization of variability, the TEMZ model requires presetting the density of high-energy electrons to vary with time in a “red noise” form to reproduce the observed variability results. 
One limitation of these models is the necessity for a predetermined variation pattern of physical parameters within the radiation region to generate red noise variability characteristics. The assumed variation pattern typically aligns with the characteristics of red noise. In our model, the variation of the underlying physical parameters is not predetermined. 
However, the radiation pulse shape of the blobs is preset and has not been obtained by rigorously solving the particle acceleration and radiation cooling within the blobs under the magnetic reconnection (or perhaps shock) process. This limits the ability of the model to predict the time-delay correlation of the inter-band light curves. 
Nevertheless, the PSD simulation results of the asymmetric pulse (see Section IV B) and the ability of the model to roughly reproduce the observed FVA characteristics in blazars strongly suggest that the specific particle acceleration and radiative cooling processes within the blob may not be the key factors in shaping the long-term light curve characteristics of blazars. 

Webb \textit{et al.}~\cite{webb2021nature,webb2023structure} applied shock turbulence models to analyze the micro-variability phenomenon of blazars, resolving each pulse in the micro-variability structure to reveal the scales of turbulent cells and their plasma properties \citep[also see][]{bhatta201372}. 
Our model and the shock turbulence model both attribute the stochastic variability of blazars to the superposition of pulses from different turbulent blobs, although the focus here is primarily on the formation of long-term stochastic variability characteristics in blazars. 
From the simulation results in Figure 3(b), the specific triggering mechanism of a single turbulent blob pulse (which determines the specific pulse profile) has a small influence on the overall PSD shape, but it is still notable that in the shock turbulence model, the blob pulses have a consistent KRM pulse structure, which could influence the shape of the PSD in the high-frequency region. 

Recently, \citet{liu2023multizone} proposed a multi-zone stochastic dissipative model for blazar jets to explain the low-state radiation properties of blazars. The model assumes that there are numerous radiation zones (spots) in the jet, generated by stochastic dissipation events. The probability of dissipation events occurring at a distance $r$ in the jet, $p(r)$, is parameterized as a power-law form $p(r) = A r^{-\alpha}$, where $A$ is a normalization constant and $\alpha$ is the index describing the probability distribution. Within a given timescale $T$, the expected total number of spots $N$ in the jet can be obtained by integrating $p(r)$ over the jet's span. The location and radiation parameters of each spot are randomly generated using the Monte Carlo method, and the light curve is ultimately formed by superimposing them. It can be noted that this model shares significant similarities with our model in terms of the main implementation, but there are differences in the physical interpretation and focus. 
Their model mainly focuses on the rigorous radiative process solutions for each blob to explain the low-state radiation properties and polarization variations of blazar jets. However, they do not explore the origin of the PSD characteristics of the long-term stochastic variability in blazars. 
In contrast, the model presented here primarily starts from a statistical perspective, focusing on how the stochastic model explains the PSD and variability amplitude characteristics. The aim is to uncover the most fundamental physical factors that determine the long-term variability characteristics of blazars. 

\section{Conclusion}
This paper presents a novel minimal physical model to elucidate the long-term random variability in blazars. The model is built upon the turbulent cascade process occurring within magnetized plasma jets, where energy is radiated with highly anisotropy at small-scale structures. The essence of long-term random variability is the result of the superposition of numerous discrete flaring events. Remarkably, under the premise that all underlying physical parameters are naturally uniformly random, this model spontaneously generates the observed red noise variability characteristics in blazars. 
The main advantages of this model are: 1) The model parameters are simple and have clear physical meanings, without the need to predefine the form of underlying parameter fluctuations. The red noise characteristics of variability spontaneously form, and the model exhibits self-similarity at multiple time scales, naturally explaining the universal PSD structure exhibited by different types of blazars. 2) The model predicts that when the cascade process results in a relatively flat energy distribution of blobs, the spectral index of the PSD in the high-frequency region will be steeper compared to the spectral index under the DRW model. This is consistent with recent observational findings in the variability of AGNs, providing a possible theoretical explanation. 
The model is also capable of reproducing the observed features of the FVA in blazars, indicating that the specific particle acceleration and radiative cooling processes within the blob may not be the key factor shaping the long-term stochastic variability of blazars. 
However, the model still has some limitations.
The radiation pulse profile of each blob in the current model is preset and does not strictly solve for the pulse structure formed by the particle acceleration and radiative cooling processes under the magnetic reconnection (or perhaps shock) process within the blob. This limitation prevents the model from predicting the time-delay correlations of light curves between different bands. 

\begin{acknowledgments}
We sincerely thank the anonymous referee for constructive suggestions. We are grateful for the financial support from the National Natural Science Foundation of China (No. 12103022) and the Special Basic Cooperative Research Programs of Yunnan Provincial Undergraduate Universities Association (No. 202101BA070001-043 and 202301BA070001-104). N.D. is sincerely grateful for the financial support of the Xingdian Talents Support Program, Yunnan Province (NO. XDYC-QNRC-2022-0613). 
\end{acknowledgments}
\bibliography{apssamp}
\end{document}